\documentclass[twoside,12pt,amsmath,amssymb]{article}
\usepackage{epsfig}
\usepackage{graphicx,epsf}
\usepackage{amsfonts}
\usepackage{amsmath}
\usepackage{amssymb}
\usepackage{mathrsfs}

\newcommand{\be}{\begin{eqnarray}}
\newcommand{\ee}{\end{eqnarray}}
\newcommand{\bea}{\begin{eqnarray}}
\newcommand{\eea}{\end{eqnarray}}

\begin{document}
\date{September 17, 2009}
\title{ \vspace{1cm} Explaining LSND and MiniBooNE using altered neutrino dispersion relations}
\author{Sebastian Hollenberg, Octavian Micu\footnote{Speaker.}~, Heinrich P\"as\\
Technische Universit\"at Dortmund, Dortmund, Germany}
\maketitle
\begin{abstract}
We investigate the possibility to explain the MiniBooNE anomaly by CPT and Lorentz symmetry violating neutrino-antineutrino oscillations in a two generation framework. We work with four non-zero CPT-violating parameters that
allow for resonant enhancements in neutrino-antineutrino oscillation phenomena in vacuo which are suitably described in terms of charge conjugation eigenstates of the system. We study the relation between the flavor, charge conjugation and mass eigenbasis of neutrino-antineutrino oscillations and examine the interplay between the available CPT-violating parameter space and possible resonance structures.
\end{abstract}
The data from the MiniBooNE collaboration \cite{AguilarArevalo:2008rc, AguilarArevalo:2009xn} reveal a resonance-like excess of events in the low-energy neutrino channel, but do not show a deviation from the expected oscillation pattern in the antineutrino channel. The LSND collaboration \cite{Aguilar:2001ty} on the other hand observes an excess in the antineutrino channel. Also, there is a hint that the MiniBooNE results for the resonance-like anomaly in the $\nu_{\mu}$ data actually look more like a $\nu_{\mu} \to \bar{\nu}_e$ conversion than $\nu_{\mu} \to \nu_e$ events.
In order to understand these yet unexplained anomalies, neutrino oscillation scenarios with altered dispersion relations have recently received attention. The resonance structure might possibly indicate new physics and motivated different possible explanations
of this phenomenon \cite{Pas:2005rb, Hollenberg:2009ws, Hollenberg:2009bq}. Amongst other things CPT and Lorentz symmetry violating neutrino oscillations have been proposed
\cite{Kostelecky:2003xn, Barger:2007dc}. In the framework of a CPT- and Lorentz-violating Standard Model extension with renormalizable
operators only \cite{Kostelecky:2000mm, Colladay:1998fq, Kostelecky:2003cr}, mixing between light neutrinos and antineutrinos is encountered which might provide a viable candidate when it comes to explaining resonance
features in $\nu_e \rightleftharpoons \nu_{\mu}$ and $\bar{\nu}_e \rightleftharpoons \bar{\nu}_{\mu}$ oscillation experiments.
\par
Starting from the generalized Dirac equation introduced in \cite{Kostelecky:2003cr}, we consider a model for only the first two neutrino and antineutrino generations, and allow for Lorentz- and CPT-violating interactions. The off diagonal part of the effective Hamiltonian for this case reads
\be
h_{\text{eff}} =
        \begin{pmatrix} -\frac{\Delta m^2}{4E}\cos2\theta - \frac{c_{ee}E}{2} &
                          \frac{\Delta m^2}{4E}\sin2\theta  &
                          \frac{b_{e}E}{2} & 0 \\
                          \frac{\Delta m^2}{4E}\sin2\theta  &
                          \frac{\Delta m^2}{4E}\cos2\theta - \frac{c_{\mu\mu}E}{2} &
                          0 & \frac{b_{\mu}E}{2} \\
                          \frac{b_{e}E}{2} & 0 &
                          -\frac{\Delta m^2}{4E}\cos2\theta - \frac{c_{ee}E}{2} &
                          \frac{\Delta m^2}{4E}\sin2\theta  \\
                          0 & \frac{b_{\mu}E}{2} &
                          \frac{\Delta m^2}{4E}\sin2\theta  &
                          \frac{\Delta m^2}{4E}\cos2\theta - \frac{c_{\mu\mu}E}{2}\\
           \end{pmatrix} \label{heff},
    \ee
in the basis $(\nu_e, \nu_{\mu}, \nu_e^c, \nu_{\mu}^c)$, and where $c_{ee}$ and $c_{\mu\mu}$ are Lorentz violating parameters; while $b_e$ and $b_{\mu}$ are both Lorentz- and CPT-violating parameters. The presence of the CPT-violating parameters induces a mixing between the neutrino and antineutrino sectors making neutrino-antineutrino oscillations possible.
\par
The effective Hamiltonian $h_{\text{eff}}$ can be brought to a block-diagonal form $\tilde{h}_{\text{eff}}$ with the help of a unitary matrix $U$
    \be
        h_{\text{eff}} = U~\tilde{h}_{\text{eff}}~U^{\dagger}.
    \ee
It is then convenient to change from flavor basis using the unitary matrix $U$ into a new basis
    \be
        \left(\begin{array}{c} \nu_e \\ \nu_{\mu} \\ \nu_e^c \\ \nu_{\mu}^c \end{array}\right) \to
        \frac{1}{\sqrt{2}} \begin{pmatrix} 1 & 0 & -1 & 0 \\ 0 & 1 & 0 & -1 \\ 1 & 0 & 1 & 0 \\
        0 & 1 & 0 & 1 \end{pmatrix}
        \left(\begin{array}{c} \nu_e \\ \nu_{\mu} \\ \nu_e^c \\ \nu_{\mu}^c \end{array}\right)
        = \frac{1}{\sqrt{2}} \left(\begin{array}{c} \nu_e^{} - \nu_e^c \\ \nu_{\mu}^{} - \nu_{\mu}^c \\
        \nu_e^{} + \nu_e^c \\ \nu_{\mu}^{} + \nu_{\mu}^c \end{array}\right)=\left(\begin{array}{c} \nu_{e}^- \\ \nu_{\mu}^- \\\nu_{e}^+ \\ \nu_{\mu}^+\end{array}\right),
    \ee
\begin{figure}
\centering
\raisebox{4.5cm}{$\sin^22\theta_{\text{eff}}$}
\includegraphics[scale=0.95]{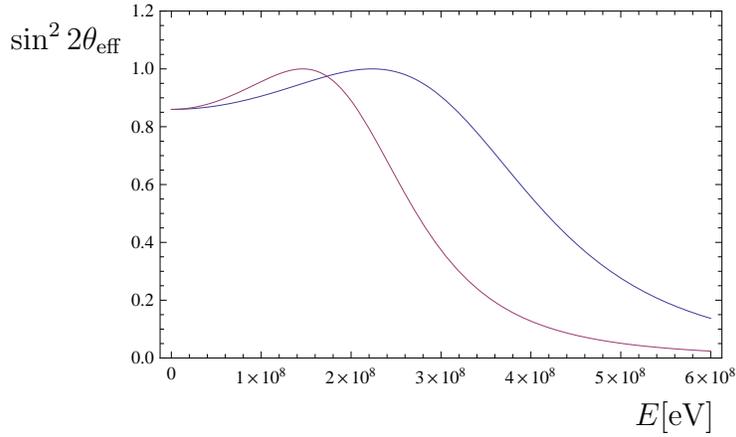}
\\{\hspace{8cm}$E[\text{eV}]$}
\caption{Resonance structures between charge conjugation eigenstates. Shown is the sine-squared of the effective mixing angles $\theta_{\mathscr{C}\text{-odd}}$ (blue curve) and
$\theta_{\mathscr{C}\text{-even}}$ (red curve). We choose $b_e=1\times 10^{-21}$, $b_{\mu}=0.6\times 10^{-21}$, $c_{\mu\mu}=3\times 10^{-21}$, $c_{ee}=2\times 10^{-21}$ for illustrative purposes. We take $\Delta m^2 = 8 \times 10^{-5} \text{eV}^2$ as well as $\sin^22\theta = 0.86$. In this case the resonance energy for the $\mathscr{C}\text{-odd}$ mixing is higher as compared to the resonance energy of the $\mathscr{C}\text{-even}$ mixing.} \label{sintheta}
\end{figure}
with the states $\nu^-$ and $\nu^+$ of the new basis being eigenstates of the charge conjugation operator $\mathscr{C}$. In the new basis the Hamiltonian is block diagonal, and the $\mathscr{C}$-even and the $\mathscr{C}$-odd sectors can be diagonalized further separately.
\par Using the usual diagonalization procedure for each sector, the effective mixing angles $\theta_{\mathscr{C}\text{-odd}}$ and $\theta_{\mathscr{C}\text{-even}}$ can be calculated independently \cite{Hollenberg:2009tr}. Depending on the values of the Lorentz- and CPT- violating parameters resonances in one or in both sectors occur. The resonance energies $E^{\mathscr{C}\text{-odd}}_{\text{res}}$ and $E^{\mathscr{C}\text{-even}}_{\text{res}}$ can also be calculated. In general the two resonance energies are different. As can be seen from Fig. \ref{sintheta}, for that particular choice of parameters both resonances exist, and the resonance energies are different. The graph also shows that in the low energy regime the effective mixing angles go to the standard mixing, which is the same for neutrinos and antineutrinos. The effective mixing angles go through maximal mixing at energies equal to the resonance energy, then as the energy increases beyond the resonance energy, the mixing goes to zero. In the limit in which the Lorentz- and CPT-violating parameters vanish the effective mixing angles equal the standard mixing angle, and the resonances disappear.
\par The effective mixing angles and resonance energies are calculated in the charge conjugation operator eigenbasis. The translation between the mass basis and the flavor basis is done via a matrix $V$ which is the product of the unitary matrix $U$ and the matrix which diagonalizes the block diagonal Hamiltonian defined with the charge conjugation eigenbasis
\be
        \left(\begin{array}{c} \nu_e \\ \nu_{\mu} \\ \nu_e^c \\ \nu_{\mu}^c \end{array}\right) =
        \frac{1}{\sqrt{2}} \begin{pmatrix} \cos\theta_{\mathscr{C}\text{-odd}} & \sin\theta_{\mathscr{C}\text{-odd}} &
        \cos\theta_{\mathscr{C}\text{-even}} & \sin\theta_{\mathscr{C}\text{-even}} \\ -\sin\theta_{\mathscr{C}\text{-odd}} &
        \cos\theta_{\mathscr{C}\text{-odd}} & -\sin\theta_{\mathscr{C}\text{-even}} & \cos\theta_{\mathscr{C}\text{-even}}
        \\ -\cos\theta_{\mathscr{C}\text{-odd}} & -\sin\theta_{\mathscr{C}\text{-odd}} & \cos\theta_{\mathscr{C}\text{-even}} &
        \sin\theta_{\mathscr{C}\text{-even}} \\
        \sin\theta_{\mathscr{C}\text{-odd}} & -\cos\theta_{\mathscr{C}\text{-odd}} & -\sin\theta_{\mathscr{C}\text{-even}} &
        \cos\theta_{\mathscr{C}\text{-even}} \end{pmatrix}
        \left(\begin{array}{c} \nu_1 \\ \nu_2 \\ \nu_3 \\ \nu_4 \end{array}\right).
    \ee
The probability of oscillation is
\be
        P(\beta \to \alpha) = \left|\sum_i V_{\beta i} e^{-iE_it} \left(V^{\dagger}\right)_{i\alpha}\right|^2 ,\label{oscprob}
    \ee
where $E_i$ are the effective energy eigenvalues of the associated Hamiltonian $h_{\text{eff}}$; and $\alpha$ and $\beta$ stand for the four different neutrino species involved, i.e. $\alpha,~\beta = \nu_e,~\nu_{\mu},~\nu_e^c,~\nu_{\mu}^c$. It is important to emphasize that the CPT-violating parameters make oscillations between neutrinos and antineutrinos from the same generation or of a different generation possible.
\par
\par
We find that even a simple choice of non-zero CPT-violating coefficients provides a workable model in which neutrino-antineutrino oscillations become possible.
The model for neutrino-antineutrino oscillations under consideration in a CPT-violating framework gives rise to new vacuum resonances \cite{Barger:2000iv} which are suitably described in terms of $\mathscr{C}$-even and $\mathscr{C}$-odd states. Resonant mixing as defined occurs between $\mathscr{C}$-flavor eigenstates rather than between common flavor eigenstates. Depending on the parameter space of the CPT-violating coefficients it is possible to have none, one (for the $\mathscr{C}$-even or $\mathscr{C}$-odd states) or two resonances (one resonance in each sector not necessarily at the same energy). These resonances are related to the mixing of $\mathscr{C}$-flavor eigenstates. Another point to be made is that at least one of the neutrino-antineutrino resonances reveals a narrower resonance width as compared to neutrino-neutrino oscillations with altered dispersion relations in the CPT conserving case. CPT violation distinguishes particles and antiparticles such that resonance peaks for neutrinos and antineutrinos are not necessarily identical. Such a behavior might be suggested by a recent analysis of the experimental neutrino data \cite{Karagiorgi:2009nb} along with the hint that the signal observed at MiniBooNE looks more like a
$\nu_{\mu} \to \bar{\nu}_e$ conversion than $\nu_{\mu} \to \nu_e$ events.
Without going into further details we mention that depending on the choice of parameters, the model predicts interesting daily and seasonal variations of neutrino oscillation observables, which result from the Earth's motion with respect to a preferred frame implied by a Lorentz-violating background field.
A detailed analysis of both the direction dependence of CPT-violating coefficients as well as the flavor oscillation probability will possibly shed light on neutrino oscillation anomalies such as LSND and MiniBooNE.

\end{document}